\documentclass[aps,prx,letterpaper,floatfix,  preprintnumbers,  reprint,
amsmath,amssymb,  showpacs,showkeys,  superscriptaddress,]{revtex4}%
\usepackage{lineno,xcolor}
\usepackage{graphicx}
\usepackage{bm}
\usepackage[breaklinks=true]{hyperref}
\usepackage{enumitem}
\usepackage{epstopdf}
\usepackage{amsmath}
\usepackage{amsfonts}
\usepackage{amssymb}%
\providecommand{\U}[1]{\protect\rule{.1in}{.1in}}
\hypersetup{bookmarksnumbered, pdfpagemode=UseOutlines,
pdfauthor={A.\ Aqeel},
pdftitle={Spin-Hall magnetoresistance and spin Seebeck effect in multiferroic CoCr$_2$O$_4$ films}, pdfdisplaydoctitle,
colorlinks=true, citecolor=blue, filecolor=blue, linkcolor=blue, urlcolor=blue}

\begin{document}
\date{\today}
\title{Spin-Hall magnetoresistance and spin Seebeck effect in spin-spiral and
paramagnetic phases of multiferroic CoCr$_{2}$O$_{4}$ films}
\author{A.\ \surname{Aqeel}}
\affiliation{Zernike Institute for Advanced Materials, University of Groningen, Nijenborgh
4, 9747 AG Groningen, The Netherlands}
\author{N.\ \surname{Vlietstra}}
\affiliation{Zernike Institute for Advanced Materials, University of Groningen, Nijenborgh
4, 9747 AG Groningen, The Netherlands}
\author{J.\ A.\ \surname{Heuver}}
\affiliation{Zernike Institute for Advanced Materials, University of Groningen, Nijenborgh
4, 9747 AG Groningen, The Netherlands}
\author{G.\ E.\ W.\ \surname{Bauer}}
\affiliation{Institute for Materials Research and WPI-AIMR, Tohoku University, Sendai,
Miyagi 980-8577, Japan}
\affiliation{Kavli Institute of NanoScience, Delft University of Technology, Lorentzweg 1,
2628 CJ Delft, The Netherlands}
\author{B.\ \surname{Noheda}}
\affiliation{Zernike Institute for Advanced Materials, University of Groningen, Nijenborgh
4, 9747 AG Groningen, The Netherlands}
\author{B.\ J.\ \surname{van Wees}}
\affiliation{Zernike Institute for Advanced Materials, University of Groningen, Nijenborgh
4, 9747 AG Groningen, The Netherlands}
\author{T.\ T.\ M.\ \surname{Palstra}}
\email{[e-mail: ]t.t.m.palstra@rug.nl}
\affiliation{Zernike Institute for Advanced Materials, University of Groningen, Nijenborgh
4, 9747 AG Groningen, The Netherlands}

\begin{abstract}
We report on the spin-Hall magnetoresistance (SMR) and spin Seebeck effect
(SSE) in multiferroic CoCr$_{2}$O$_{4}$ (CCO) spinel thin films with Pt
contacts. We observe a large enhancement of both signals below the spin-spiral
($T_{s}=28\,\mathrm{K}$) and the spin lock-in transitions
($T_{\mathrm{lock-in}}=14\,\mathrm{K}$). The SMR and SSE response in the spin
lock-in phase are one order of magnitude larger than those observed at the
ferrimagnetic transition temperature ($T_{c}=94\,\mathrm{K}$), which indicates
that the interaction between spins at the Pt$|$CCO interface is more efficient
in the non-collinear magnetic state below $T_{s}$ and $T_{\mathrm{lock-in}}$.
At $T>T_{c}$, magnetic field-induced SMR and SSE signals are observed, which
can be explained by a high interface susceptibility. Our results show that the
spin transport at the Pt$|$CCO interface is sensitive to the magnetic phases
but cannot be explained solely by the bulk magnetization.

\end{abstract}
\maketitle

\renewcommand\linenumberfont{\normalfont\tiny\sffamily\color{gray}}

\section{Introduction}

Ferro (ferri) magnetic insulators (FMI) with normal metallic (NM) contacts
that support the spin Hall effect (SHE) and its inverse (ISHE) open new
functionalities in the field of spintronics. The SHE refers to a charge
current that induces a transverse spin current, which can be injected to 
actuate a metallic or insulating ferromagnet. The ISHE converts a
spin current pumped out of a ferromagnet into a transverse charge current in
the normal metal. These concepts have been confirmed by many experiments on
FMI$|$NM bilayers of a magnetic insulator (usually yttrium iron garnet) and a
heavy normal metal (usually platinum), for example spin-pumping by
ferromagnetic resonance \cite{Sandweg2011,Castel2012,Hahn2013,Harii2011}, the
spin Seebeck effect
(SSE)~\cite{Uchida_nmat_2010,Schreier2013,Aqeel2014,Wu2015}, the spin Peltier
effect~\cite{Flipse2014} and the spin Hall magnetoresistance
(SMR)~\cite{Althammer2013,Vlietstra2013,Nakayama2013,Vlietstra2013_Appl,Hahn2013_PRB,Isasa2014}%
. In the SMR, both SHE and the ISHE act in a concerted manner to allow
electrical detection of the FMI magnetization direction. The SSE refers to the
conversion of thermal excitations of the magnetic order parameter (spin waves
or magnons) into a spin current pumped into NM and detected by the ISHE. The
SSE and the SMR have been investigated up to now only for a limited number of
insulating ferrimagnets (garnets and spinels) with collinear magnetizations
and recently, in an antiferromagnetic insulator~\cite{Han2014}. However,
magnetic insulators come in a large variety of magnetic order. Especially
fascinating are non-collinear magnets with competing magnetic interactions
(spin frustration) induced by competing next-nearest neighbor exchange.
Alternatively, the spin-orbit coupling, such as the Moriya-Dzyaloshinskii
interaction favor complex spiral configurations and skyrmion order. While the
coupling of non-collinear magnetizations with spin, charge and heat transport
is currently one of the hottest subjects in magnetism, its role in the SMR and
SSE appears to have not been studied yet.

Non-collinear magnetism emerges when the second nearest neighbor magnetic
interactions are of the same order as the first one, generating geometrical
frustration that favors spin canting. Various spin-spiral orders, like proper
screw, cycloidal, longitudinal-conical and transverse-conical spiral, have
been observed. The cycloidal and the transverse-conical spiral orders break
the inversion symmetry and induce a spontaneous electrical polarization,
making these spiral magnetic systems
multiferroic~\cite{Fiebig2005,Cheong2007,Tokura2010,Ramesh2007,Kimura2008}.
Here, we focus on non-collinear and multiferroic CoCr$_{2}$O$_{4}$ (CCO) thin
films, reporting to the best of our knowledge first experimental observation
of the SMR and the SSE in the Pt$|$CCO system for a wide range of temperatures
including the ferrimagnetic and the spin-spiral phases.

\begin{figure}[ptb]
\centering
\includegraphics[width=0.7\textwidth,natwidth=310,natheight=342]{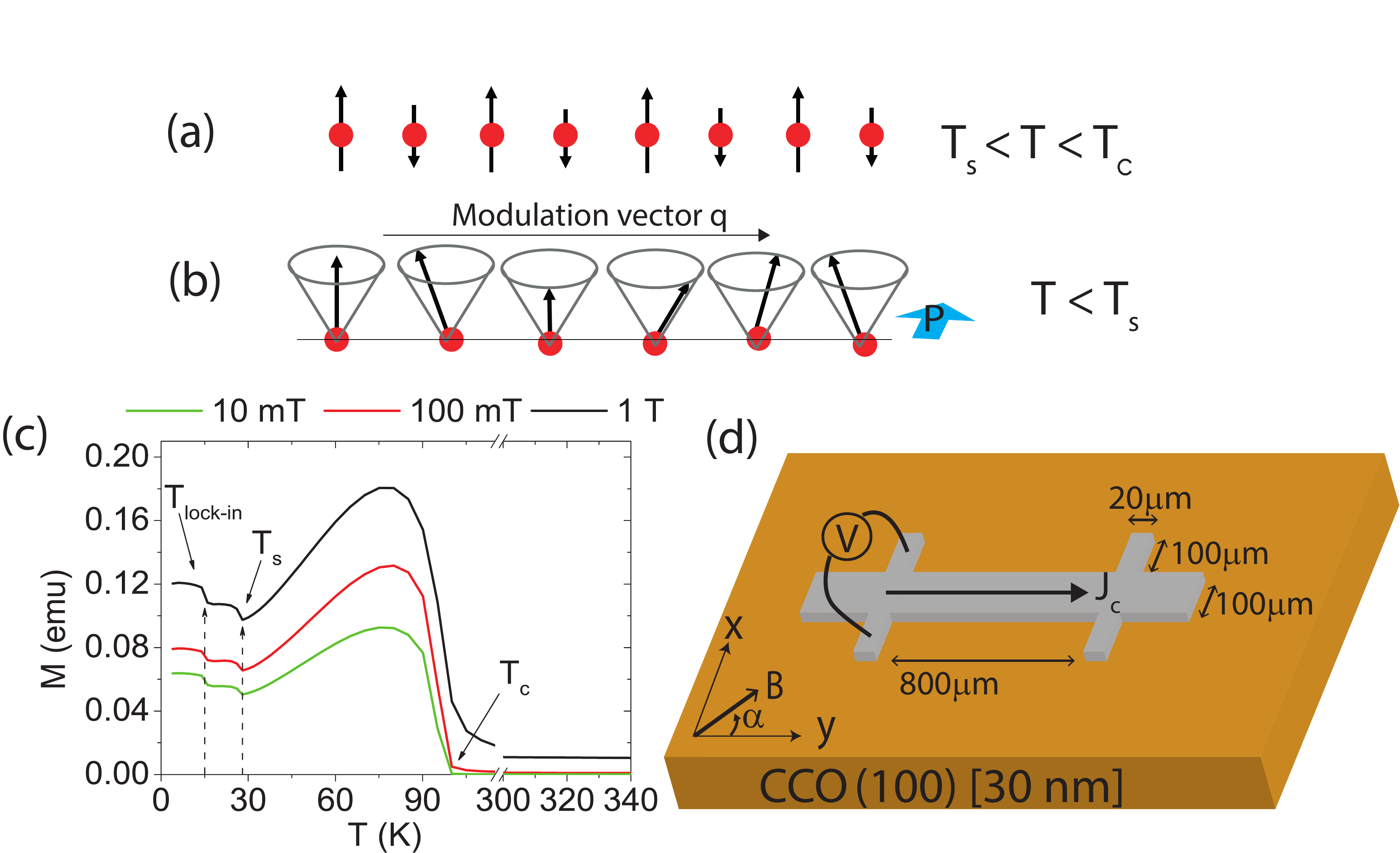}\caption{(a,b) Impression
of the types of magnetic order in CCO. (a) Ferrimagnetic state for
$T_{s}<T<T_{c}$ and (b) transverse conical spiral state for $T<T_{s}$ with an
electrical polarization P. Here, $T_{c}$ and $T_{s}$ are the Curie transition
temperature to a collinear ferrimagnetic phase and that the spin-spiral
transition temperature, respectively. (c) The temperature dependence of the
zero-field cooled magnetization of the CCO target powder used for the film
deposition for different applied magnetic fields. $T_{\mathrm{lock-in}}$ is
the spin lock-in magnetic transition temperature. (d) Device configuration for
the transverse resistance (planar Hall effect) measurement of the Pt film on
top of CCO. }%
\label{fig:1}%
\end{figure}

CCO is one of the rare multiferroic materials with linear magnetoelectric
coupling~\cite{Yamasaki2006,Choi2009}. It has a normal spinel structure with
three sublattices. The Co$^{2+}$ ions are located exclusively in the
tetrahedral sites (forming one sublattice) while Cr$^{3+}$ ions reside in the
octahedral sites (in two sublattices). The Cr$^{3+}$ ions form a pyrochlore
lattice with magnetic geometrical frustration~\cite{Tomiyasu2004,Bordacs2009}.
Bulk CCO exhibits long-range collinear ferrimagnetic order below
$T_{c}=93\,K-95\,K$~\cite{Yang2012,Tomiyasu2004,Lawes2006}, with an easy axis
of magnetization along the [001]-direction, as illustrated in Fig.~\ref{fig:1}%
(a) and (c). At $T_{s}=28\,K$, the ferrimagnetic long-range order adopts an
additional short-range spiral order as illustrated in Fig.~\ref{fig:1}(b) and
(c), which is known as the spin-spiral phase transition. The conical spiral
has 48$^{\text{o}}$, 71$^{\text{o}}$ and 28$^{\text{o}}$ cone angles with the
[001]-direction for the Co-, Cr I- and Cr II- sublattices, respectively. Below
$T_{s}$, magnetic spin-spirals move the oxygen atoms off-center due to the
inverse Dzyaloshinskii-Moriya interaction~\cite{Singh2011,Yang2012}, which
results in the appearance of a spontaneous electrical polarization. At
$T_{\mathrm{lock-in}}=14\,K$, the spin-spiral becomes commensurate with the
lattice by the spin-lattice coupling, which is known as the spin lock-in
transition. 

Here, we report a systematic study of the SMR and the SSE in the Pt$|$CCO
bilayers from low to room temperature ($T=5\,\mathrm{K}-300\,\mathrm{K}$). At
each temperature, we record the dependence on the angle of an in-plane applied
magnetic field. We observe strong effects of the CCO magnetic order with
largest signals in the spin lock-in phase at $T<14\,\mathrm{K}$.

\section{Sample growth and characteristics}

$30\,\mathrm{nm}$-thick CCO films were grown on (001) MgO substrates by pulsed
laser deposition (PLD). Sintered ceramic CCO targets \cite{Mufti2010} were
ablated with a KrF excimer laser light with a wavelength of $248\,\mathrm{nm}$
and a repetition frequency of $0.5\,\text{Hz}$. During the deposition, the MgO
substrates are kept at $500\,^{\text{o}}\mathrm{C}$ in oxygen plasma
atmosphere, having a base pressure of $0.01\,\text{mbar}$. Afterwards, the
films were cooled down, with $5\,^{\text{o}}\mathrm{C}$ per minute in 0.5 bar
O$_{2}$ atmosphere. Before further device fabrication, the films were annealed
at $200\,^{\text{o}}\mathrm{C}$ for $60\,\mathrm{min}$ in an O$_{2}$
atmosphere. The crystal structure of the CCO films was determined by x-ray
diffraction (XRD), where the rocking curves show high crystalline quality with
a Full Width at Half Maximum (FWHM) below $0.03^{\text{o}}$.

The magnetization of the CCO films was measured by a SQUID magnetometer. The
films show an in-plane magnetic anisotropy with a coercive field $H_{c}$
around 2\thinspace T. For temperatures below $50\,\mathrm{K}$, the magnetic
transitions are hard to detect due to a large paramagnetic substrate
background compared to the small magnetic moment of the thin film (see
Appendix~\ref{S3}). Nevertheless, SQUID
measurements~\cite{ChoiKang2009,Liu2014} and X-ray resonance magnetic
scattering both indicate the same magnetic transitions in thin films as
reported for bulk CCO. In order to demonstrate the magnetic transitions in
CCO, we measured the temperature dependence of the magnetization of bulk CCO
targets used for the film deposition. Fig.~\ref{fig:1}(c) shows a
magnetization of the CCO targets with the same transitions as reported in
literature~\cite{Tsurkan2013}. We carried out transport experiments on
Hall-bar structures patterned by electron beam lithography onto which a
$4\,\mathrm{nm}$-thick Pt film was deposited by dc-sputtering, as shown in
Fig.~\ref{fig:1}(d).

\section{Measurement techniques}

SMR and SSE have been measured simultaneously by lock-in detection
\cite{Vlietstra2014}. Using two Stanford SR-830 Lock-in amplifiers, the first
and second harmonic voltage response were recorded separately. To minimize the
background voltage, we used the transverse instead of the longitudinal
configuration, as schematically shown in Fig.\ref{fig:1}(d), which is also
referred to as \textquotedblleft planar Hall\textquotedblright\ voltage. The
SMR signals scale linearly with the applied current and therefore detected in
the Pt resistance of the first harmonic response~\cite{Vlietstra2014}. The
current-induced SSE scales quadratically with the applied current and
therefore detected in the second harmonic response. The angular dependence of
the SMR and the SSE were studied by rotating an external magnetic field in the
plane of the film~\cite{Schreier_APL_2013} that above $T_{c}$ induces a
magnetization or (if large enough) aligns the direction of the CCO
magnetization below $T_{c}$. The in-plane angle $\alpha$ of the magnetic field
is defined relative to the current direction along the $y$-axis as indicated
in Fig.~\ref{fig:1}(d). All experiments were carried out in a quantum design
PPMS system, at magnetic fields $B\geq2\,\text{T}$ and for temperatures from
$5\,$to $300\,\mathrm{K}$.

\section{Results and discussion}

\subsection{Spin-Hall magnetoresistance}

\begin{figure}[ptb]
\includegraphics[width=0.8\textwidth,natwidth=310,natheight=342]{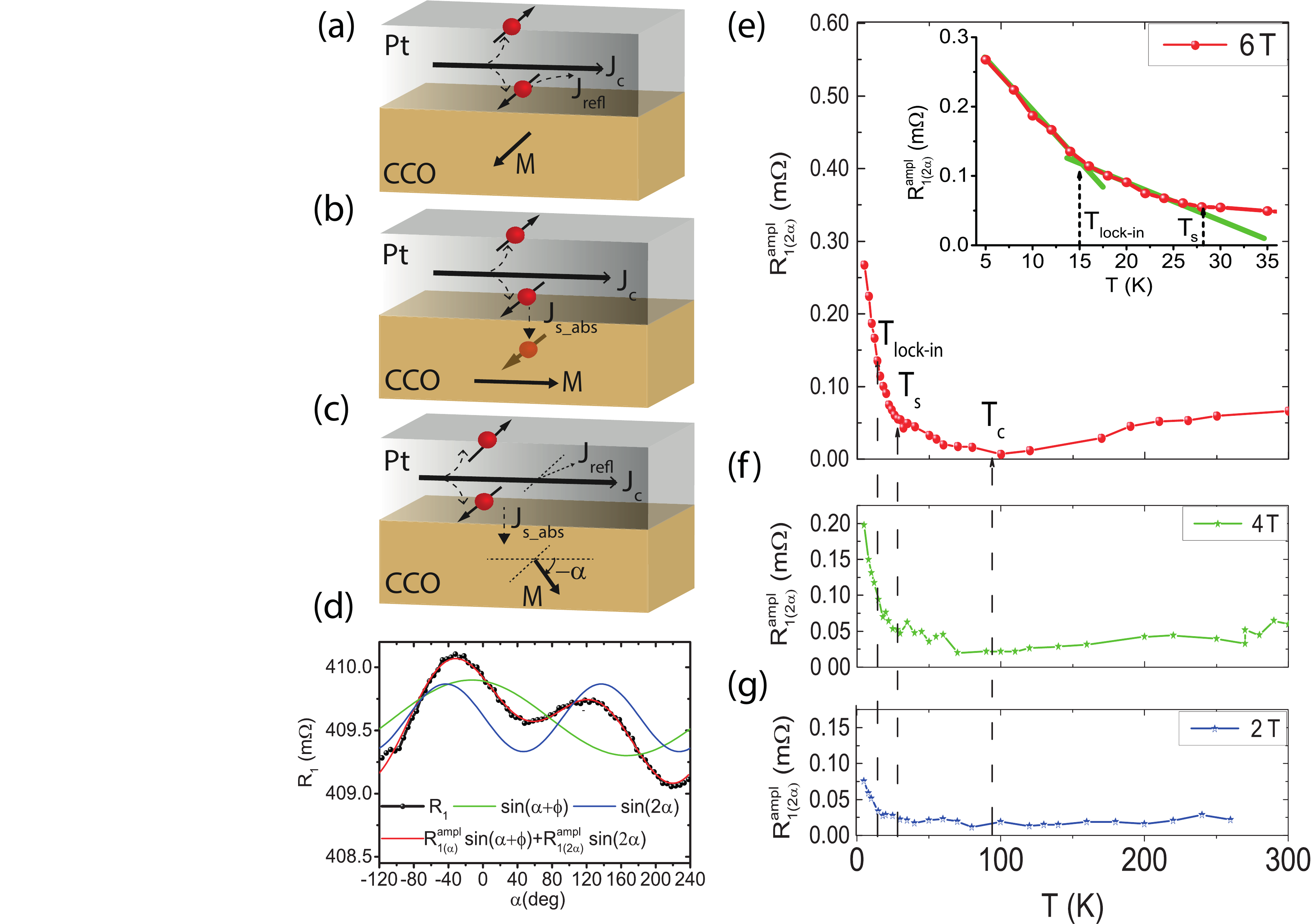}\caption{(a-c)
Illustration of the SMR in Pt$|$CCO bilayers. A charge current $I$ induces a
spin current and thereby spin accumulation at the Pt$|$CCO interface by virtue
of the SHE. (a) This spin accumulation is absorbed as a spin transfer torque
when the magnetization $\vec{M}$ is perpendicular to the current-induced spin
polarization in Pt,. (b) When $\vec{M}$ is parallel to the spin accumulation,
the spin current is reflected back into Pt, where it generates an additional
charge current, $J_{\mathrm{refl}}$ by the ISHE. (c) When $\vec{M}$ is at an
angle $\alpha$ to $J_{c}$, the component of the spin accumulation
perpendicular to $\vec{M}$ is absorbed and the component parallel are
reflected, leading to an extra charge current component normal to $J_{c}$ that
is detected in the Hall configuration (d) The angular dependence of the
1$^{\mathrm{st}}$ harmonic response in the transverse configuration,
$R_{1}=V_{1}/I$, for $I=2\,$mA at 5~K in an applied magnetic field of 6~T. The
$\sin(\alpha+\phi)$ and $\sin2\alpha$ curves illustrate the additive
contributions from the ordinary Hall effect and the SMR. (e), (f) and (g) show
the temperature dependence of the SMR, $R_{1(2\alpha)}^{\mathrm{ampl}}$ for
$J_{c}=2\,$mA at 6~T, 4~T and 2~T, respectively. Here, $R_{1(2\alpha
)}^{\mathrm{ampl}}$ is the amplitude of the $\sin2\alpha$ component from the
SMR. $T_{c}$, $T_{s}$ and $T_{\mathrm{lock-in}}$ are the ferrimagnetic,
spin-spiral and spin lock-in magnetic transition temperatures, respectively.
The inset in (e) is a zoomed-in image of the SMR signal below $T_{s}$ and
$T_{\mathrm{lock-in}}$ at 6~T. }%
\label{fig:2}%
\end{figure}

Owing to the SHE, an ac-current through the Pt Hall-bar creates an ac spin
accumulation at the Pt$|$CCO interface that can be partially absorbed or fully
reflected, depending on the interface magnetization $\vec{M}$ of the FMI (see
Fig. ~\ref{fig:2}(a) and (b)). The reflected spin currents generate an extra
charge current via the ISHE, thereby reducing the resistance. While the
longitudinal resistance establishes the maximum modulation between these two
configurations with $\alpha=0^{\text{o}}$ or $90^{\text{o}}$, the planar Hall
effect vanishes. In contrast, when $\alpha=45^{\text{o}}$, the ISHE is maximal
\cite{Chen2013}, as sketched in Fig.~\ref{fig:2}(c). The additional emf scales
with the applied current and is therefore, detected by the first harmonic
transverse resistance $R_{1}$~\cite{Vlietstra2014}, defined as $V_{1}/I$,
where $V_{1}$ is first harmonic signal of the lock-in amplifier generated by
the applied ac current amplitude $I$ (see Appendix~\ref{S1} for more details).

The $\alpha$ dependence of the first harmonic response of $R_{1}$ at 5~K in
Fig.~\ref{fig:2}(d) shows on top of the expected SMR an ordinary Hall effect
(OHE) generated by a magnetic field component normal to the film due to a
slight misalignment of the sample by an angle $\beta$. The OHE voltage has a
$\sin\left(  \alpha+\phi\right)  $ angular dependence, where the phase $\phi$
is governed by the sample tilt direction. A prefactor of $0.7\,\mathrm{\mu}%
$\textrm{V} at a current of $2\,$mA in a magnetic filed 6~T corresponds to a
tilt of $\beta<2^{\text{o}}.$ The ordinary Hall voltage of Pt$|$CCO is nearly
temperature independent and scales linearly with the applied current and
magnetic field, as expected. After subtracting the OHE from $R_{1}$, the
anticipated $\sin2\alpha$ dependence associated to the SMR remains
\cite{Althammer2013}, as illustrated in Fig.~\ref{fig:2}(d). The SMR signal,
$R_{1(2\alpha)}^{\mathrm{ampl}}$, is defined as the amplitude of the
$\sin2\alpha$ component and plotted in Fig.~\ref{fig:2}(e), (f) and (g) as a
function of temperature in magnetic fields of 6~T, 4~T and 2~T, respectively.
Exchanging the current and voltage probes in the Hall bar in filed of 6~T
leads to identical SMR profiles, confirming that the interface magnetization
is parallel to the applied field. The SMR\ of CCO and yttrium iron garnet
(YIG) are compared in Appendix~\ref{S2}.

\begin{figure}[ptbh]
\includegraphics[width=1\textwidth,natwidth=310,natheight=342]{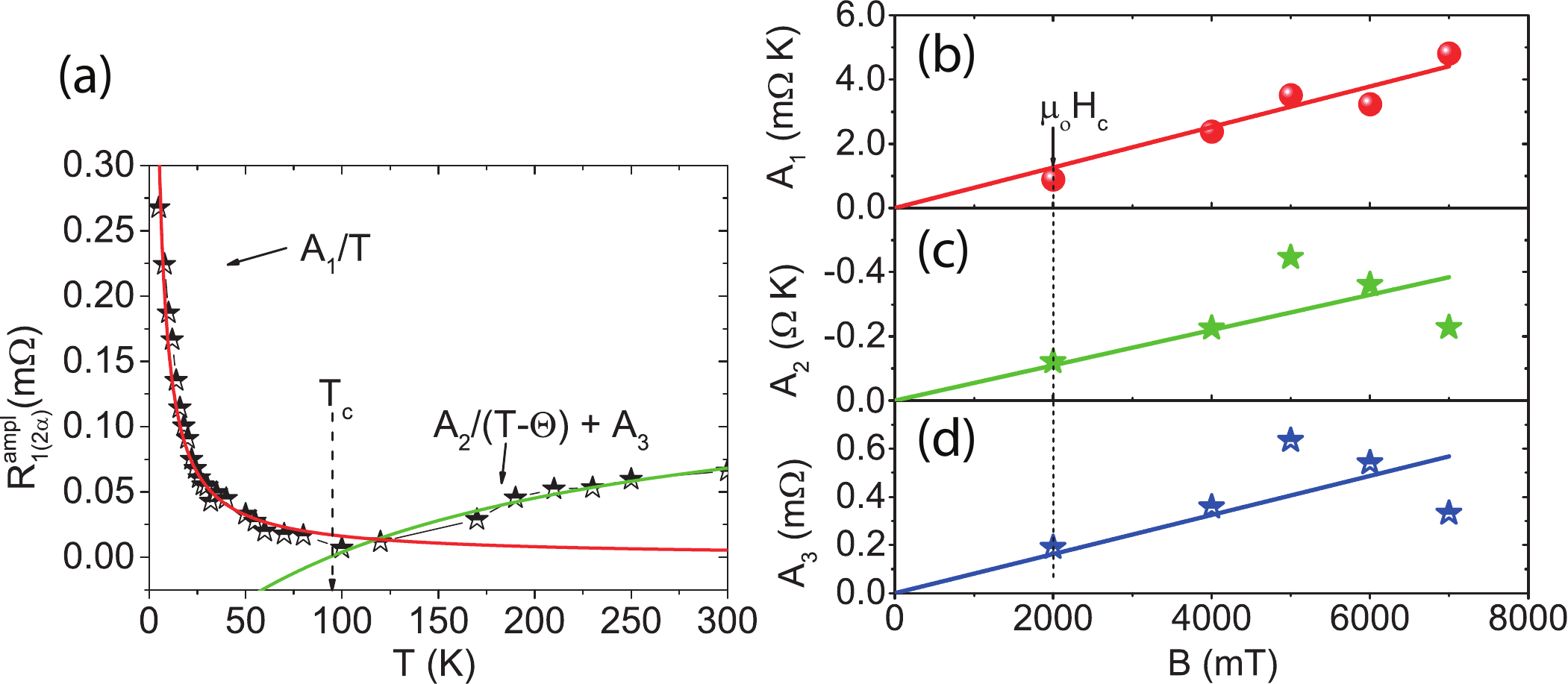}\caption{(a) shows the
temperature dependence of $R_{1(2\alpha)}^{\text{ampl}}$ in a magnetic field
of 6~T for $I=2\,mA$. The red curve shows the fit with Curie law at low
temperatures below T$_{c}$ and the green curve shows the fit at the high
temperature range above T$_{c}$ with Curie-Weiss law (for $\theta
_{CW}=-550\,K$). (b), (c) and (d) show the field dependence of fitting
parameters $A_{1}$, $A_{2}$ and $A_{3}$, respectively. }%
\label{fig:3}%
\end{figure}

Below the collinear ferrimagnetic transition temperature $T_{c}=94\,\mathrm{K}%
$, the SMR signal increases with decreasing temperature. The inset of
Fig.~\ref{fig:2}(e) recorded at a magnetic field of 6~T shows a distinct
change of slope at the spin-spiral transition temperature to a non-collinear
magnetic phase, $T_{s}=28\,\mathrm{K}$. At $T<T_{s}$, the SMR signal is more
than one order of magnitude larger than the signal observed at $T_{c}$. A
further decrease in temperature below the spin lock-in transition temperature,
$T_{\mathrm{lock-in}}=14\,\mathrm{K}$, doubles the SMR compared to $T_{s}$.
This observation indicates that the exchange interaction between metal and
magnet in the Pt$|$CCO system is more efficient in the non-collinear spiral
phase than in the collinear ferrimagnetic phase. The maximum SMR signal is
observed in the spin lock-in phase, when the period of the spin spiral becomes
commensurate with the lattice. Below $T_{c}$, $R_{1}=A_{1}T^{-1}$ (see
Fig.~\ref{fig:3}(a)) gives an excellent fit, where $A_{1}$ scales linearly
with the applied magnetic field, as shown in Fig.~\ref{fig:3}(b). $A_{1}$
should proportional to the interface spin-mixing conductance that vanishes
with interface magnetization. Above $T_{c},$ all magnetization is generated by
the applied magnetic field and $A_{1}$ is a measure of the interface
paramagnetic susceptibility. For a magnetic interface, we anticipate a
bilinear $A_{1}(B),$ with a large slope at low magnetic fields that reflects
the expulsion of magnetic domain walls. The extrapolation of the high-field
data should lead to a finite cut-off at zero magnetic fields (as in studies of
the anomalous Hall effect). However, the extrapolation of $A_{1}(B)$ does not
lead to a statistically significant $A_{1}(0)$. At present we therefore cannot
confirm whether the observed SMR signals reflect the paramagnetic
susceptibility of a non-magnetic interface or a spontaneous interface
magnetization texture.

In order to shed more light on the $T$-dependence for $T<T_{c}$, we compare
the SMR\ with the bulk magnetization, as shown in Fig.~\ref{fig:1}(c). The
$T_{c}$, $T_{s}$ and $T_{\mathrm{lock-in}}$ transition temperatures
established from the SMR, closely corresponds to the transition temperatures
observed in the bulk magnetization. In the spin-spiral and spin lock-in
phases, the bulk magnetization does not depend much on temperature, as shown
in Fig.~\ref{fig:1}(c), in stark contrast to the SMR below $T_{s}$ (Fig.
~\ref{fig:2}(e)). The SMR therefore does not directly reflect the CCO bulk
magnetization. Theoretically, the SMR arises from the modulation of the spin
current at the NM$|$FMI interface by the magnetization \cite{Chen2013} and is
roughly proportional to the density of oriented magnet moments at the
interface \cite{JiaSTT2011}. The SMR enhancement below $T_{s}$ thus reflects
an increased order of the magnetic moments at the interface with reduced
temperature. The moments at the interface can be either Co ions, contributing
to the ferrimagnetic component, or Cr ions responsible for the cycloidal
component of the spin-spiral. The ferrimagnetic component of magnetization
saturates around $28\,\mathrm{K}$, therefore cannot explain the enhanced SMR
below $28\,\mathrm{K}$. However, neutron scattering experiments show that the
spiral component below $T_{s}=28\,\mathrm{K}$ strongly depends on
temperature~\cite{Tomiyasu2004}, similar to the SMR signal. We therefore
venture that the development of the spiral order should be explained by a
strongly temperature-dependent ordering of Cr ions that cannot be observed in
the global magnetization. This implies that by a simple transport experiment
we can distinguish ferrimagnetic from cycloidal components of the interface magnetization.

We observe a finite SMR in the paramagnetic phase for $T>T_{c}$
complementing reports on spin pumping~\cite{Shiomi2014} and the spin Seebeck
effect\cite{Wu2015} in the paramagnetic state. In CCO the magnetic
susceptibility, for $T>T_{c}$, follows a Curie-Weiss law with a negative
Curie-Weiss temperature $\theta_{CW}=-550\,\mathrm{K,}$ which is evidence for
antiferromagnetic correlations. The high ratio $\left\vert \theta
_{CW}\right\vert /T_{c}\approx6$ indicates a significant magnetic frustration
due to competing sublattice exchange interactions in CCO~\cite{Tsurkan2013},
resulting in short range order above $T_{c}$. The SMR signal above $T_{c}$
increases with temperature until it saturates to a constant value around
room-temperature, as shown in Fig.~\ref{fig:2}(e-g). The SMR signal for
$T>T_{c}$ provides evidence for an unusual interface magnetic susceptibility
of our films. We fitting the SMR signals with the Curie-Weiss law, but a
single fitting parameter $A_{2}$ does not capture the contributions from the
molecular fields. Much unlike the bulk magnetic susceptibility, the SMR signal
is suppressed at $T_{c}$ which can be taking care of by introducing an
additional parameter $A_{3}$, as shown in Fig.~\ref{fig:3}(a). Both fitting
parameters $A_{2}$ and $A_{3}$ scale linearly with the applied magnetic field,
as shown in Fig.~\ref{fig:3}(c) and (d). This results support our conclusion
drawn earlier that the SMR signal cannot be solely explained by the bulk
magnetization of CCO, even in the paramagnetic phase. The
temperature-independent background modelled by $A_{3}$ and the negative sign
of $A_{2}$ remain unexplained by indicate that the short-range order reported
for bulk CCO above $T_{c}$ is importantly modified at the interface to Pt.

\subsection{Spin Seebeck effect}

\begin{figure}[ptb]
\includegraphics[width=0.7\textwidth,natwidth=310,natheight=342]{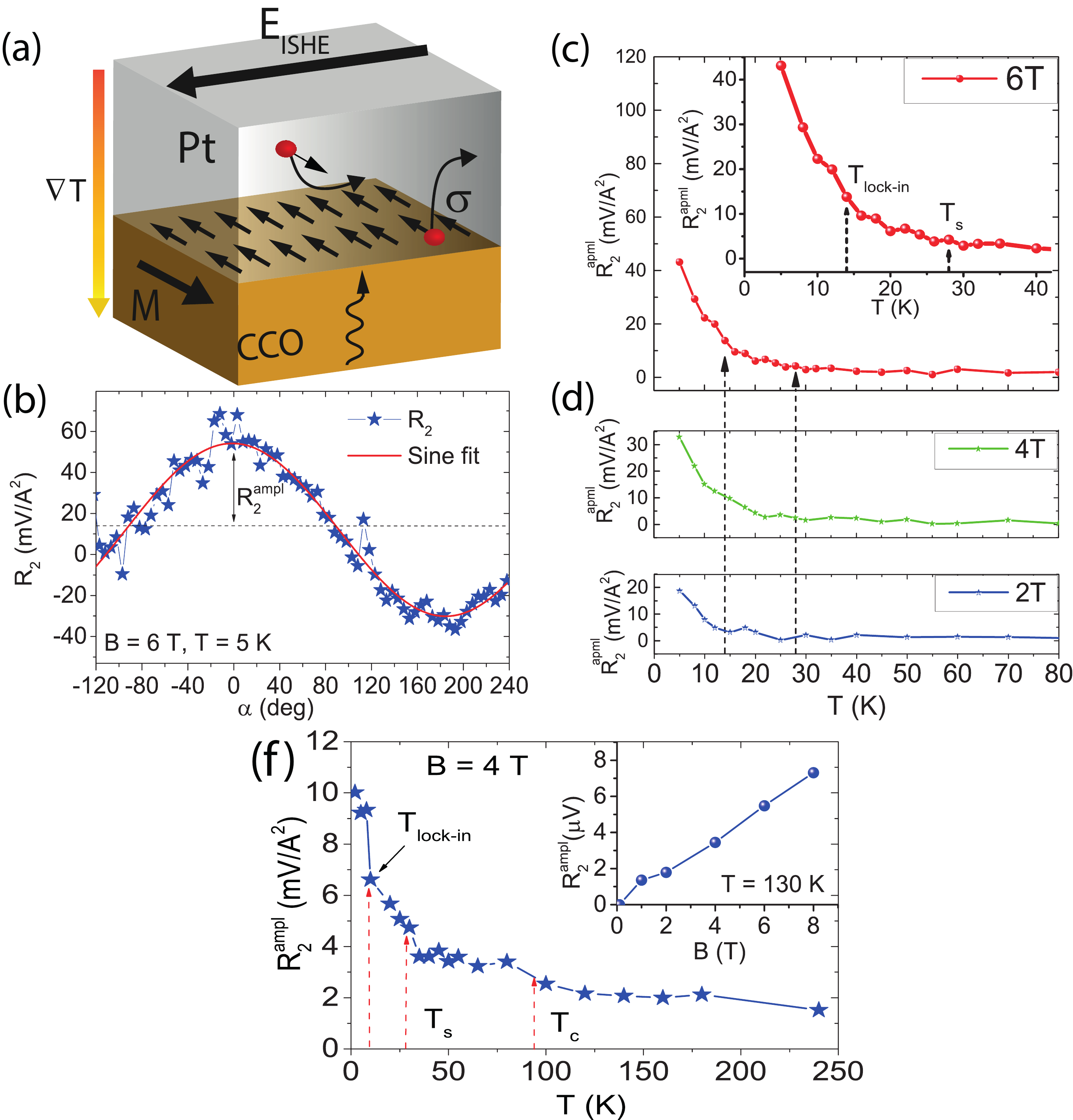}\caption{(a) Spin Seebeck
effect in Pt$|$CCO. A thermal gradient creates magnons at the interface by
absorption of a spin current from Pt with polarization $\vec{\sigma}\Vert
\vec{M}$ , where $\vec{M}$ is the CCO interface magnetization$.$ The spin
current generates an electromotive force $\vec{E}_{\text{ISHE}}$ by the ISHE.
(b) The angular dependence of the second harmonic response, $R_{2}=\sqrt
{2}V_{2}/I^{2}$ at $5\,K$, for $I=2\,\mathrm{mA}$ in an applied magnetic field
of 6~T. (c), (d) and (e) show the temperature dependence of $R_{2}%
^{\mathrm{ampl}}$ at 6~T, 4~T and 2~T for $I=2\,\mathrm{mA}$, respectively.
The inset of (c) emphasizes the enhancement of the SSE signal below the spin
spiral ($T_{s}$) and spin lock-in ($T_{\mathrm{lock-in}}$) transition
temperatures in a magnetic field of 6~T. (f) Temperature dependence of
$R_{2}^{\mathrm{ampl}}$ in the non-linear current regime ($I~=~5~\mathrm{mA}%
$), with inset showing the magnetic field dependence of $R_{2}^{\mathrm{ampl}%
}$ at $T=130\,\mathrm{K}$. }%
\label{fig:4}%
\end{figure}

We now discuss the SSE signal observed in the second harmonic response. It is
caused by the Joule heating of the Pt Hall-bar, generating a heat current into
the ferromagnet which is absorbed by magnons. This heat current is associated
with a spin current polarized along the magnetization direction, which can be
detected electrically by the ISHE, as sketched in Fig.~\ref{fig:4}(a). The
second harmonic response $R_{2}=\sqrt{2}V_{2}/I^{2}$, where $V_{2}$ is the
second harmonic signal of the lock-in amplifier at a phase set at $\phi
=-90^{\text{o}}$~\cite{Vlietstra2014}. The observed second harmonic response of the
transverse resistance obeys the $\sin\alpha$ angular dependence anticipated
for the SSE, as shown in Fig.~\ref{fig:4}(b) We define the amplitude of the
$R_{2}^{\mathrm{ampl}}=R_{2}/\sin\alpha$ for each temperature and magnetic
field strength. The SSE signal for Pt$|$CCO has the same sign as for Pt$|$YIG
(see Appendix~\ref{S2}). Fig.~\ref{fig:4}(c), (d) and (e) show the temperature
dependence of $R_{2}^{\mathrm{ampl}}$ in magnetic fields of 6~T, 4~T and 2~T,
respectively. 

In the collinear ferrimagnetic state at $T<T_{c}$, the SSE response increases
with decreasing temperature. Fig.~\ref{fig:4}(c) gives evidence of a large SSE
enhancement below the spin spiral transition temperature $T_{s}$ and again
below $T_{\mathrm{lock-in}}$. The SSE signal below $T_{s}$ is five times
larger than the signal observed at $T_{c}$. The inset of Fig.~\ref{fig:4}(c)
illustrates that the SSE at $T<T_{\mathrm{lock-in}}$ increases by an order of
magnitude from $T=80\,\mathrm{K}$. Moreover below $T_{s}=28\,\mathrm{K}$, the
SSE signal scales linearly with the applied magnetic field. The SSE is too
noisy to provide as clear evidence for phase transitions at $T_{c}$, $T_{s}$
and $T_{\mathrm{lock-in}}$ as the SMR signals do.

Conventional thermoelectric effects become small with decreasing temperatures,
so what causes the observed remarkable enhancement at low temperatures? We can
understand the temperature dependence of the SSE signal by considering the
contributions from different magnetic sublattices
\cite{Ohnuma2013,Geprags2014}. CCO is a collinear ferrimagnet above $T_{s}$,
with three sublattices associated with Co and oppositely polarized Cr~I and
Cr~II ion moments \cite{Tomiyasu2004}, but without any magnetization
compensation below $T_{c}$. The magnetic sublattices contribute to the SSE by
correlated thermal fluctuations. The coupled sublattices have acoustic
(ferromagnetic) and optical (antiferromagnetic) modes. The fundamental
ferromagnetic modes govern the low energy excitations that are probed by
ferromagnetic resonance and Brillouin light scattering. The optical modes
are below $T_{c}$ shifted to higher energy by the exchange interaction. In the
collinear ferrimagnetic state close to $T_{c}$, the optical modes are still
significantly occupied but the acoustic modes are slightly dominant. The
acoustic and optical modes contribute to the SSE with different sign and
therefore cancel to a large extent. By decreasing the temperature, the
exchange splitting of the optical modes increases and therefore, become
increasingly depleted. The suppression of the thermal pumping of the optical
modes therefore, leads to an apparent enhancement of the spin Seebeck effect at
lower temperature. A similar mechanism explains the low temperature sign
change of the SSE of the ferrimagnetic insulator Gd$_{3}$Fe$_{5}$O$_{5}$
(GdIG)~\cite{Geprags2014} and can be used to understand the temperature
dependence of the SSE in YIG at temperatures above 300~K~\cite{Uchida2014}.

In contrast to collinear magnetic order, the magnetization texture of a spin
spiral is sensitively modulated by an external magnetic field
~\cite{Kamenskyi2013}. This might contribute as well to the observed magnetic
field dependence of the SSE below $T_{s}$, as shown in Fig.~\ref{fig:4}(c-e).
The corresponding increase in the SSE with applied magnetic field below
$T_{s}$ may persist until the Co$^{2+}$ and Cr$^{3+}$ momenta of $3\mu
_{B}/\text{ion}$ are fully aligned. No signs of magnetization saturation were
observed at magnetic fields up to 30 T~\cite{Tsurkan2013}, so it would be very
interesting to find out whether also the spin Seebeck effect can be further
enhanced in such high magnetic fields.

The observed correlation between the SMR and SSE provides evidence that the
spin-mixing conductance, i.e., the transport measure of the exchange
interaction between ferromagnet and normal metal, does play an important role.
Other factors, such as the magnetization damping and the magnon transport to
the interface that is affected by the magnon-phonon
interaction~\cite{Torgashev2012}, may contribute to the observed enhancement
in the SSE below $T_{s}$. A quantitative description of the SSE can be
approached by atomistic spin simulations that take the full spin wave spectrum
associated to the three sublattices with different cone angles, chirality, and
damping parameters, as well the spin mixing conductances of the interface to
Pt into account.

Above $T_{c}$ the SSE cannot be established at small heating current levels
($I=2~\mathrm{mA}$), as shown in Fig.~\ref{fig:4}(c). However, at a larger
current ($I=5\,\mathrm{mA}$) in a magnetic field of 4~T, a finite SSE signal
is detected above $T_{c}$ and up to room temperature, see Fig.~\ref{fig:4}(f).
At these current levels, $T_{\mathrm{lock-in}}$ is shifted to lower
temperature by $4\,\mathrm{K}$ due to sample heating, and non-linearities kick
in, i.e., the SSE voltage $\nsim I^{2}$. The SSE still scales linearly with
the applied magnetic field as shown for $T=130\,\mathrm{K}$ in the inset of
Fig.~\ref{fig:4}(f). The presence of the SSE above $T_{c}$ can be explained by
the large longitudinal magnetic susceptibility of CCO films that is
responsible for the SMR response as well ~\cite{Shiomi2014}.

In summary, by lock-in detection we simultaneously measured the SMR and SSE in
Pt$|$CCO bilayers. The temperature dependence of the SMR and, though less so,
the SSE, exposes distinct anomalies at the magnetic phase transitions. A
remarkable enhancement of both SMR and SSE signals is observed at low
temperatures ($T<T_{s}$). The SMR is more than one order of magnitude larger
at $T<T_{\mathrm{lock-in}}$ as compared to the signals around $T_{c}$. We
relate the observed enhancement of the SMR below $T_{s}$ to contributions from
the cycloidal spiral projected onto the spin accumulation at the Pt$|$CCO
interface. The SSE signal also increases by a factor of two when lowering
temperature below $T_{s}$. The temperature dependence of the SSE does not
simply reflect the bulk magnetization; instead the intricate magnetization
dynamics of coupled sublattices needs to be considered. Our results suggest
that the magnons from the complicated Cr-sublattice magnetization texture
plays a essential role in the SSE. Vice versa, we establish that the SMR and
SSE are powerful instruments that complement ferromagnetic resonance and
neutron scattering techniques to analyze the magnetization dynamics of complex
oxides including multiferroics.

\begin{acknowledgments}
We would like to acknowledge
J. Baas, H. Bonder, M. de Roosz and J. G. Holstein for technical assistance.
This work is supported by the Foundation for Fundamental Research on Matter
(FOM), NanoNextNL, a micro- and nanotechnology consortium of the government
of the Netherlands and 130 partners, by NanoLab NL, InSpin EU-FP7-ICT Grant No 612759 and the
Zernike Institute for Advanced Materials National Research Combination,
Grants-in-Aid for Scientific Research (Grant Nos. 25247056, 25220910,
26103006\U{ff09}, and DFG Priority Programme 1538\ (BA 2954/2).
\end{acknowledgments}


\appendix

\section{CCO film magnetization}

\label{S3} \begin{figure}[ptbh]
\includegraphics[width=0.5\textwidth,natwidth=310,natheight=342]{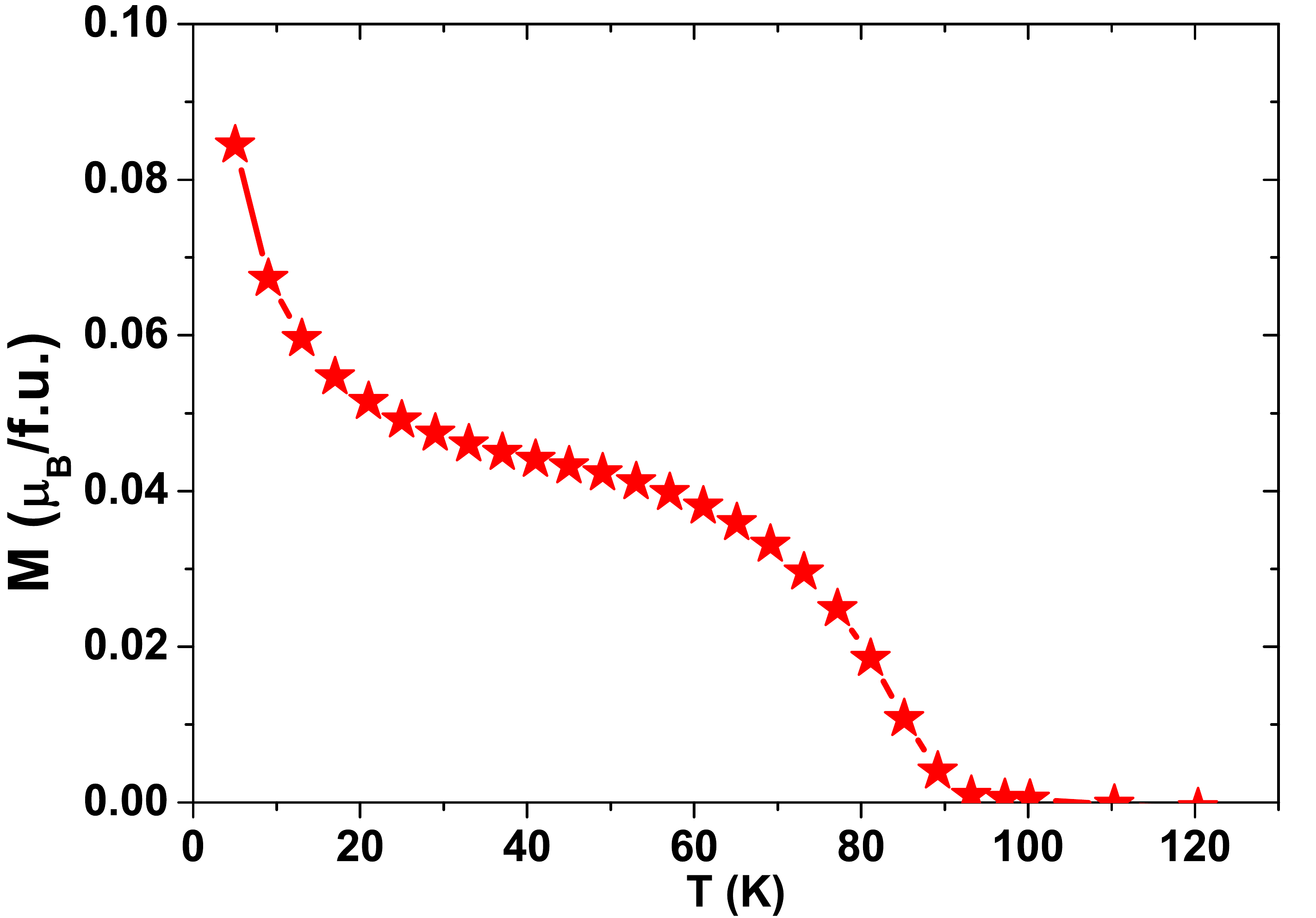}\caption{In-plane
magnetization of a CCO film on a MgO substrate in a magnetic field of 0.01~T
after cooling in 1~T. }%
\label{fig:6}%
\end{figure}
The temperature-dependent magnetization of the CCO film in
Fig.~\ref{fig:6} as measured by SQUID magnetometry is evidence for a phase
transition at $T_{c}=94\,\mathrm{K}$ to collinear ferrimagnet order, but the
spin-spiral and spin lock-in transitions are not visible, because of the small
magnetization of a thin film as explained in the main text. The apparent
increase in the magnetization below 20~K is probably caused by paramagnetic
impurities in the substrate.

\section{Lock-in detection}

\label{S1}

All measurements shown in the main text are carried out using a lock-in
detection technique~\cite{Vlietstra2014} with $I\leq5\,\mathrm{mA}$ ac current
bias in the Pt film. The generated voltage can be expanded as
\begin{equation}
V(t)=R_{1}I(t)+R_{2}I^{2}(t)+R_{3}I^{3}(t)+\cdots, \label{eq1}%
\end{equation}
where $R_{n}$ represent the $n$-th harmonic response. For $I(t)=\sqrt{2}%
I_{o}\sin\omega t$, with angular frequency $\omega$ and rms value $I_{0}$, the
harmonic response coefficients $R_{n}$ are obtained by measuring the different
frequency components $\left(  1\omega,2\omega,\cdots\right)  $ by a lock-in
amplifier. The detected $n$-th harmonic response at a set phase $\phi$ can be
written as
\begin{equation}
V_{n}(t)=\frac{\sqrt{2}}{T}\int_{t-T}^{t}\sin\left(  n\omega s+\phi\right)
V(s)ds. \label{eq2}%
\end{equation}
Focussing on first and second order response, we define the output voltage of
the lock-in amplifier for the first and second harmonic responses by using
Eq.~(\ref{eq1}) and Eq.~(\ref{eq2}) as:
\begin{equation}%
\begin{array}
[c]{c}%
V_{1}=I_{0}R_{1}\\
V_{2}=I_{0}^{2}R_{2}/\sqrt{2}%
\end{array}
\text{ for }%
\begin{array}
[c]{c}%
\phi=0^{\text{o}}\\
\phi=-90^{\text{o}}%
\end{array}
\label{eq34}%
\end{equation}
The SMR and the SSE signals appear in the first and second harmonic responses
in Eq.~(\ref{eq34}), respectively.

\begin{figure}[ptbh]
\includegraphics[width=0.8\textwidth,natwidth=310,natheight=342]{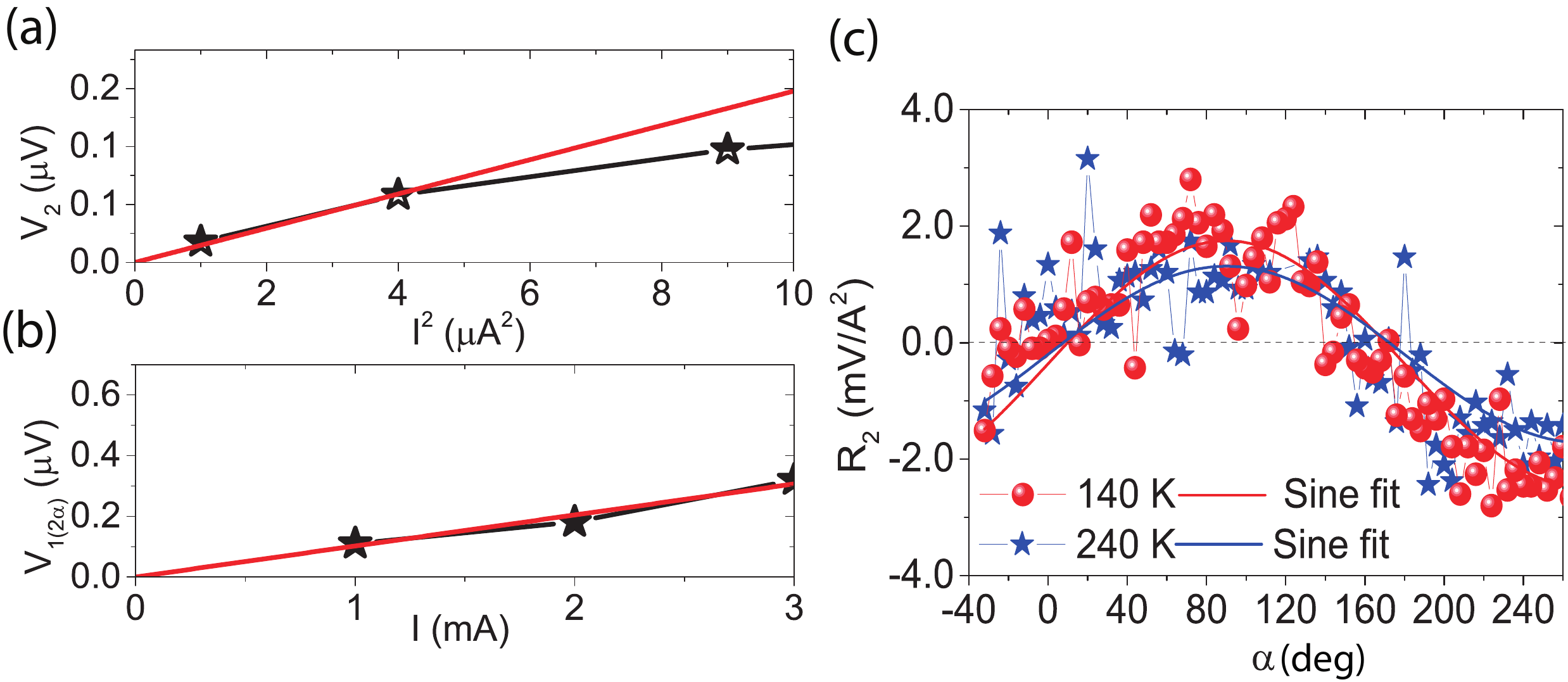}\caption{(a) Dependence
of the second harmonic response V$_{2}$ on $I^{2}$ due to the SSE, generated by current induced
heating and (b) dependence of first harmonic contribution V$_{1(2\alpha})$ ,
due to the SMR, by an ac-current $I$ sent through the Pt Hall-bar in a
magnetic field of 4~T. The SMR scales linear and the SSE scales quadratic I in
the linear-regime ($I\leq2\,mA$). (c) The angular dependence of second
harmonic response, $R_{2}$, for $I=5\,mA$, $H=4$~T at $T=140\,K$~and~$240\,K$.
}%
\label{fig:5}%
\end{figure}

In the linear response regime at currents $I\leq2\mathrm{\,mA}$, the SMR
scales linearly and the SSE scales quadratically with the applied current as
shown in Fig.~\ref{fig:5}(a,b). However, at $I>2\mathrm{\,mA}$, the SMR (SSE)
do no longer depend linearly (quadratically) on the current $I$. The results in the
main text for the SMR are gathered in the linear regime. At $T>T_{s}$ the SSE
signal decreases rapidly and we record the SSE also at the high current
$I=5\mathrm{\,mA}$, as shown in Fig.~\ref{fig:5}(c). 

\section{SMR and the SSE in Pt$|$YIG vs. Pt$|$CCO}

~\label{S2} To compare the SMR and the SSE response in Pt$|$CCO system to that
of Pt$|$YIG, the angular dependence of Pt$|$YIG system is also systematically
studied at different temperatures. A $4\,\mathrm{nm}$-thick Pt Hall-bar was
deposited on a $4\times4\,\mathrm{mm}^{2}$ YIG film by dc
sputtering~\cite{Vlietstra2014}. The Pt Hall-bar has a length of
$500\mathrm{\,\mu m}$ and a width of $50\mathrm{\,\mu m}$, with side contacts
of $10\,\mathrm{\mu m}$ width. A $200\,\mathrm{nm}$ thick YIG, single-crystal,
film is used and the film is grown by liquid phase epitaxy on a (111) Gd$_{3}%
$Ga$_{5}$O$_{12}$ (GGG) substrate.

We observe the same sign of the SMR signals for Pt$|$CCO and Pt$|$YIG, as
expected. A change of the sign of SSE has been observed in compensated
ferrimagnet~\cite{Geprags2014} but for both Pt$|$YIG and Pt$|$CCO system the
sign remains the normal for all temperatures ($5-300\,\mathrm{K}$).

\begin{figure}[ptbh]
\includegraphics[width=0.8\textwidth,natwidth=310,natheight=342]{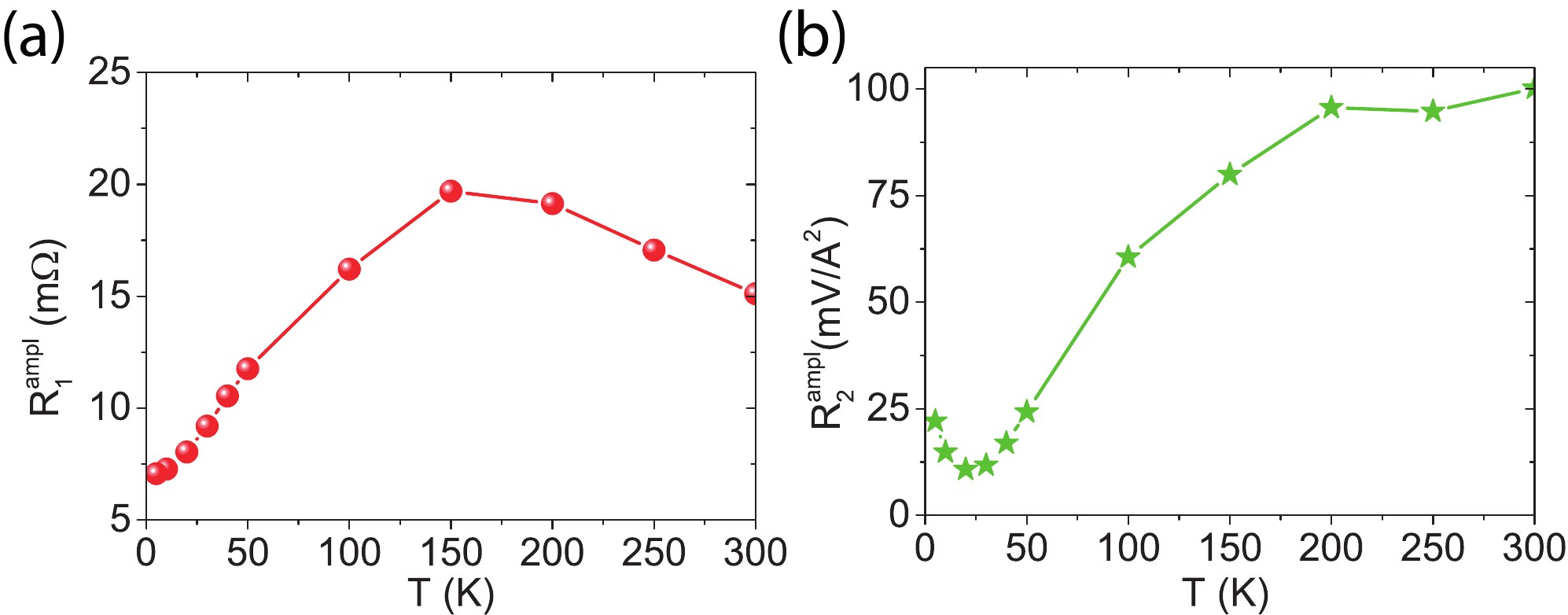}\caption{(a) Temperature
dependence of the SMR and (b) the SSE in the Pt$|$YIG system for $2.5\,mA$
current through the Pt Hall-bar in a magnetic field of 0.1~T. }%
\label{fig:7}%
\end{figure}

The temperature dependence of the SMR and SSE is observed to be very different
in the Pt$|$CCO as compared to the Pt$|$YIG system. In Pt$|$CCO, both SMR and
the SSE signals are enhanced at temperatures much lower than $T_{c}$, as shown
in Fig.~\ref{fig:2}(e) and Fig.~\ref{fig:4}(c) with maximal values around
$5\,\mathrm{K}$ in the spin lock-in phase. In contrast, Pt$|$YIG displays
conventional behavior, with both SMR and SSE larger at room temperature than
at low temperatures. Fig.~\ref{fig:7}(a) shows the temperature dependence
Pt$|$YIG SMR at relatively large current levels of $I=2.5\,\mathrm{mA}$. The
SMR signal slightly increases when decreasing temperature until 150~K and
decreases again when cooling even more~\cite{Marmion2014}. The decrease in the
SMR signal in Pt$|$YIG has been ascribed to a decrease in the Pt spin-Hall
angle with decreasing temperature \cite{Meyer2014}. The SMR signal observed at
5~K is twice smaller than that at room-temperature. The SSE in Pt$|$YIG does
not change much until $T=200\,\mathrm{K}$, while a further decrease in
temperature suppresses the SSE as shown in Fig.~\ref{fig:7}(b). A small
increase in the SSE is observed for $T<30\,\mathrm{K}$. 

The ordinary Hall effect as observed in the first harmonic response of
transverse resistance (as shown in Fig.~\ref{fig:2}(d)) 
is almost temperature independent in Pt$|$CCO system and scales
linearly with the applied field as expected. The ordinary Hall effect is not
observed in the Pt$|$YIG system because a much weaker field suffice to saturate the
YIG magnetization ($H_{c}<1\,\mathrm{mT}$). Moreover, the SMR signal observed
in the Pt$|$CCO (shown in Fig.~\ref{fig:2}(e)) is smaller than that in the
Pt$|$YIG system, resulting in a relatively larger contribution of the ordinary
Hall effect. The SMR response observed at 5~K in the Pt$|$CCO is more than one
order of magnitude smaller than the signal observed in the Pt$|$YIG.
\end{document}